# Deep photonic reservoir computer for nonlinear equalization of 16-level quadrature amplitude modulation signals

Rui-Qian Li,[1] Yi-Wei Shen,[1] Zekun Niu,[2] Guozhi Xu,[2] Jingyi Yu,[1] Xuming He,[1] Lilin Yi[2,a)] and Cheng Wang[1,3,a)]

**AFFILIATIONS**

[1]School of Information Science and Technology, ShanghaiTech University, Shanghai 201210, China
[2]State Key Lab of Advanced Optical Communication Systems and Networks, School of Electronic Information and Electrical Engineering, Shanghai Jiao Tong University, Shanghai 200240, China
[3]Shanghai Engineering Research Center of Energy Efficient and Custom AI IC, ShanghaiTech University, Shanghai 201210, China

a)Authors to whom correspondence should be addressed: wangcheng1@shanghaitech.edu.cn; lilinyi@sjtu.edu.cn

**ABSTRACT**

Photonic reservoir computer (PRC) is a kind of real-time and adaptive recurrent neural network, where only weights in the readout layer require training. PRC is a promising tool to deal with the crucial issue of nonlinear equalization in optical fiber communications. Here we theoretically show a deep PRC for the nonlinear equalization of coherent signals with the format of 16-level quadrature amplitude modulation (16-QAM). The deep PRC consists of cascading injection-locked Fabry-Perot lasers with optical feedback. Both the in-phase component and the quadrature component of the 16-QAM signals are simultaneously injected into the deep PRC in parallel, based on the wavelength multiplexing of Fabry-Perot lasers. It is demonstrated that the deep PRC exhibits strong capability for the nonlinearity compensation of coherent signals. The Q factor is improved by more than 1 dB for 16-QAM signals with launch powers above 10 dBm, associated with a bit rate of 240 Gbps and a transmission distance of 50 km.

## I. INTRODUCTION

Photonic reservoir computer (PRC) is an optical computing system based on a special type of recurrent neural networks (RNNs). Unlike traditional RNNs, weights in the input layer and the hidden layer of PRCs are randomly fixed, while only weights in the readout layer require training. Consequently, the PRCs have merits of low training complexity and fast training speed, which enable real-time and adaptive reconfiguration of the weights.[1] One implementation approach of PRCs involves interconnecting physical neurons spatially on a single chip.[2,3] However, the scalability of this scheme is usually limited, due to the lack of nonlinear neurons[4] and the transmission loss of optical waveguides.[5] Another main approach is employing one physical neuron in combination with a feedback loop to produce a large number of virtual neurons, which are interconnected temporally through intrinsic nonlinear dynamics.[6] This scheme is usually implemented by using a semiconductor laser with an optical feedback loop[7] or using an optical modulator with an optoelectronic feedback loop.[8,9] The former one is more power efficient because the feedback loop does not require any optical-to-electrical conversion.

In optical fiber communication systems, the signal is usually distorted due to the linear chromatic dispersion and the nonlinear Kerr effect of optical fibers, which in turn limits the channel capacity according to the Shannon's law.[10] The linear distortion is well mitigated by various digital signal processing (DSP) algorithms like the feedforward equalizer (FFE).[11,12] However, it remains struggles to address the nonlinear impairments, because common DSP algorithms, such as the digital backpropagation method,[13] the perturbation method,[14] and the Volterra series method,[15] have too high computational complexity and hence too large power cost.[16] In order to reduce the computational complexity, various deep neural network structures have been proposed for the fiber nonlinearity compensation,[17,18] such as the multilayer perceptron (MLP), the convolutional neural network (CNN),[19] the long short-term memory (LSTM), the gated recurrent units (GRU),[20] as well as advanced neural networks with attention mechanism.[21,22]

Owing to the inherent high-speed and low-latency nature of analog optical systems, PRCs are highly suitable for compensating for the nonlinear impairment of optical fibers. Most investigations have focused on the nonlinearity compensation of intensity-modulation direct-detection (IM-DD) optical communication systems. Argyris *et al.* experimentally demonstrated that a laser-based PRC improved the bit-error-rate (BER) of a 25 Gbps non-return-to-zero (NRZ) signal by two orders of magnitude at 45 km transmission distance.[23] Similarly, Vatin *et al.* reported that a PRC based on a vertical-cavity surface-emitting laser (VCSEL) with two polarization modes reduced the BER by about one order of magnitude, for a 25 Gbps NRZ signal at 50 km distance.[24] Sackesyn *et al.* showed that a waveguide-based PRC with 32 on-chip nodes improved the BER of a 32 Gbps NRZ signal at 25 km distance by about one order of magnitude as well.[25] In addition, Peng *et al.* theoretically reported the performance of a semiconductor optical amplifier-based PRC for the nonlinear equalization of a 28-Gbps NRZ signal at a long haul transmission of about 4000 km.[26] Estebanez *et al.* applied a laser-based PRC in the data recovery of a 56 GBaud 4-level pulse amplitude modulation (PAM-4) signal at 100 km transmission, and its performance was better than a DSP-based Kramers-Kronig receiver.[27] In comparison with IM-DD systems, coherent communication systems suffer more from the nonlinear fiber impairment, because they involve both amplitude modulation and phase modulation. Consequently, the nonlinear equalization of coherent signals is much more challenging due to the increased dimensionality of complex signals. Sorokina *et al.* theoretically proposed the application of a fiber-based PRC in the distortion mitigation of 16-, 64-, and 256-quadrature amplitude modulation (QAM) signals at 30 GBaud 100 km transmission.[28] Masaad *et al.* theoretically investigated the nonlinear equalization performance of a waveguide-based PRC in combination with a Kramers-Kronig receiver. It was found that its performance outperformed the FFE in processing a 64-QAM signal at 64 GBaud 100 km transmission.[29] Zelaci *et al.* theoretically used the waveguide-based PRC to ease the requirement of the carrier-to-signal power ratio of Kramers-Kronig receivers for the equalization of 32 GBaud 16-QAM signal at 80 km distance.[30] Very recently, Masaad *et al.* experimentally demonstrated the use of a waveguide-based PRC for the equalization of 28 GBaud 4- and 16-QAM signals at 20 km distance.[31] In addition, Boikov *et al.* theoretically presented a PRC based on evanescently coupled GaAs microrings, which was used to recover the signals of 50 GBaud 16-QAM at 20 km distance.[32]

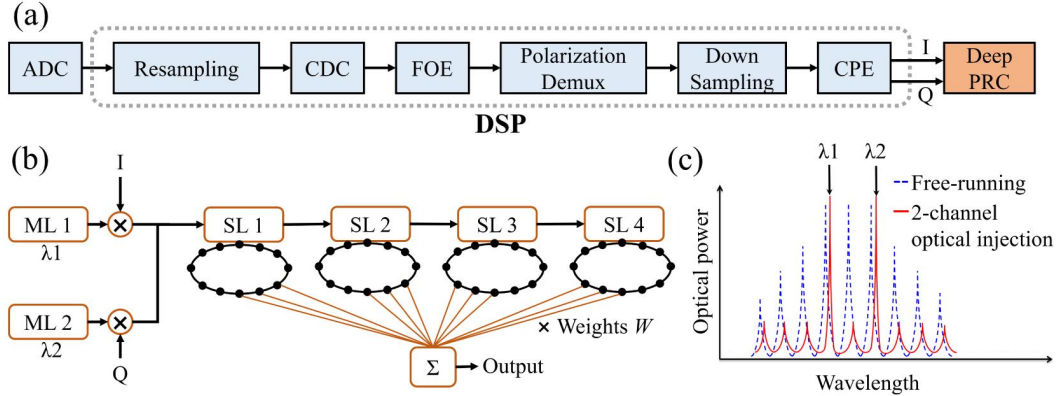

**FIG. 1.** (a) Pipeline of the signal processing with both DSP and PRC. CDC: chromatic dispersion compensation; FOE: frequency offset estimation; CPE: carrier phase estimation. (b) Schematic architecture of the deep PRC based on cascading injection-locked Fabry-Perot lasers. ML: mater laser; SL: slave laser. (c) Schematic spectrum of the Fabry-Perot laser with (solid line) and without (dashed line) optical injection.

Most PRCs used for the nonlinear equalization in literature only have a single hidden layer. However, according to the principle of deep learning, the deep architecture with multiple hidden layers can substantially leverage the representation capability of reservoir computing networks.[33,34] Our previous work proposed a deep PRC based on cascading injection-locked lasers, where the interconnection between successive hidden layers is all-optical without any optical-to-electrical conversion.[35,36] It was experimentally proved that the deep PRC exhibited strong capability of nonlinear equalization, and the BER of a 25 Gbps NRZ signal was reduced by about 90% for a transmission distance of 50 km and a high launch power of about 20 dBm. In this work, we propose to compensate for the nonlinear impairment of 16-QAM signals using the deep PRC. In this scheme, both the in-phase (I) component and the quadrature (Q) component of the 16-QAM signal are simultaneously processed in parallel, through the wavelength multiplexing of Fabry-Perot lasers. It is theoretically found that the deep PRC improves the Q factor by more than 1 dB for launch powers above 10 dBm, in association with a transmission rate of 240 Gbps (30 GBaud) and a transmission distance of 50 km. In addition, the deep PRC can compensate for the nonlinear fiber impairment up to hundreds of kilometers.

## II. ARCHITECTURE AND MODEL OF THE DEEP PRC

Here we consider a single-channel optical fiber transmission system carrying a dual-polarization (DP) 16-QAM signal. The pipeline of the signal processing at the receiver side is shown in Fig. 1(a). The received signal is amplified by an EDFA before converting to the digital domain through an analog-to-digital converter (ADC). The DSP module processes multiple functions except the nonlinear equalization. The output signal of the DSP is sent to the deep PRC module for the compensation of nonlinear fiber impairment. In the DSP module, the signal is firstly resampled to match the desired processing rate. The chromatic dispersion compensation (CDC) block compensates for the linear dispersion in the frequency domain.[37] The frequency offset estimation (FOE) block estimates and corrects the frequency difference between the carrier wave and the local oscillator, based on the algorithm of fourth power compensation.[38] Then, the x-polarization signal and the y-polarization signal are separated through the block of polarization demultiplexing (Polarization Demux). Both signals are down-sampled to reduce the data processing rate. Finally, the carrier phase estimation (CPE) block estimates and corrects the carrier phase, using the algorithm of blind phase search.[39,40,41] The output of the DSP demodulates the I and Q components for both x ($I_x$, $Q_x$) and y ($I_y$, $Q_y$) polarizations. In this work, we use the deep PRC to compensate for the nonlinearity of the x-polarization signal as an example. The deep PRC architecture in Fig. 1(b) comprises three parts: a 2-channel input layer, 4 hidden reservoir layers, and a readout layer. Every hidden layer consists of a Fabry-Perot (FP) slave laser (SL) with an optical feedback loop. All the free-running FP lasers exhibit multiple longitudinal modes as shown in Fig. 1(c) (dashed line). The optical feedback is operated in the stable regime. That is, the feedback strength is below the critical feedback level, beyond which the laser may exhibit chaotic oscillations.[12] The SL with optical feedback in each hidden layer produces a large number of virtual neurons with a temporal separation defined by the mask in the input layer.[8] All the SLs are injection-locked in series through the unidirectional optical injection technique.[36] The optical injection is operated in the stable regime as well, which is bounded by the Hopf bifurcation and the saddle-node bifurcation.[42,43] In the input layer, two individual single-mode lasers of different wavelengths $\lambda_1$ and $\lambda_2$ are used as the master lasers (MLs), respectively. As illustrated in Fig. 1(c), the wavelength of each ML is tuned to be similar to one longitudinal mode of the SLs (solid line). In such way, each ML injection locks one mode of the four SLs in the stable regime, simultaneously. All other longitudinal modes of the SLs are suppressed due to the reduction of the gain in the laser medium (see Fig 1(c)).[36,44] The two MLs act as the carrier waves of the input signals in parallel. That is, the wavelength division multiplexing (WDM) technique is used for the parallel signal processing. This parallel architecture can be easily extended to process both polarizations of the I and Q components by using four longitudinal modes of the FP laser in Fig. 1(c), which process the four components of $I_x$, $Q_x$, $I_y$, $Q_y$, respectively. Details of the parallel PRC architecture refer to our previous work in Ref. 42, 45. Both the I component and the Q component of the 16-QAM signal from the DSP are pre-processed through the masking process.[5] The pre-processed I and Q components are superimposed onto the carrier waves of the two MLs through intensity modulators, respectively. In the readout layer, the neuron states of all the four hidden layers are recorded simultaneously, and the weighted sum provides the target value of the nonlinear equalization, which is the signal without any distortion. The readout weights of the deep PRC are obtained through the algorithm of ridge regression.

The deep PRC architecture is described through the rate equation approach. Here we employ quantum dot (QD) lasers as the SLs in the reservoir layers. In comparison with quantum well lasers, QD lasers have advantages of low lasing threshold, high temperature stability, and direct epitaxial growth capability on the silicon integration platform.[46,47] In the rate equation model, we take into account the carrier dynamics in the wetting layer ($N_{RS}$), the excited state ($N_{ES}$), and the ground state ($N_{GS}$).[48] The carrier numbers of the QD laser in the *l-th* layer are described by:

$$\frac{dN_{RS}^l}{dt} = \eta \frac{I^l}{q} + \frac{N_{ES}^l}{\tau_{RS}^{ES}} - \frac{N_{RS}^l}{\tau_{ES}^{RS}}(1-\rho_{ES}) - \frac{N_{RS}^l}{\tau_{RS}^{spon}} \quad (1)$$

$$\frac{dN_{ES}^l}{dt} = \left(\frac{N_{RS}^l}{\tau_{ES}^{RS}} + \frac{N_{GS}^l}{\tau_{ES}^{GS}}\right)(1-\rho_{ES}) - \frac{N_{ES}^l}{\tau_{GS}^{ES}}(1-\rho_{GS})$$
$$- \frac{N_{ES}^l}{\tau_{RS}^{ES}} - \frac{N_{ES}^l}{\tau_{ES}^{spon}} \quad (2)$$

$$\frac{dN_{GS}^l}{dt} = \frac{N_{ES}^l}{\tau_{GS}^{ES}}(1-\rho_{GS}) - \frac{N_{GS}^l}{\tau_{ES}^{GS}}(1-\rho_{ES}) - \frac{N_{GS}^l}{\tau_{GS}^{spon}}$$
$$- \Gamma_P v_g \sum_{m=1}^{2} g_{GS}^{l,m} S^{l,m} \quad (3)$$

where $I$ is the pump current, $\eta$ is the current injection efficiency, $\rho_{ES}$, $\rho_{GS}$ are the carrier occupation probabilities in the ES and the GS, respectively. $\Gamma_P$ is the optical confinement factor, $v_g$ is the group velocity of light, $\tau_{ES}^{RS}$ is the carrier capture time, $\tau_{GS}^{ES}$ is the carrier relaxation time, $\tau_{ES}^{GS}$ and $\tau_{RS}^{ES}$ are the carrier escape times, and $\tau_{GS,ES}^{spon}$ is the spontaneous emission lifetime. Both the optical injection effect and the optical feedback effect are characterized by the Lang-Kobayashi model.[49,50] $g_{GS}^{l,m}$ is the material gain of the *m-th* longitudinal mode of the laser at the *l-th* layer, which is given by:

$$g_{GS}^{l,m} = \frac{a_{GS}(2\rho_{GS}-1)N_B/V_B}{1+\xi S^{l,m}/V_S + \beta\xi S^{l,n}/V_S} \quad (4)$$

where $a_{GS}$ is the differential gain, $N_B$ is the total number of dots, $V_B$ is the volume of the active region, and $V_S$ is the volume of the light in the laser cavity. $S^{l,m}$ and $S^{l,n}$ are photon numbers of the two modes, with the indices $m, n \in$ {I, Q}. The self-gain saturation effect is quantified by the gain compression factor $\xi$, and the cross-gain saturation effect is by the factor $\beta \times \xi$.[42] The photon dynamics and the phase dynamics of the *m-th* electric field of the QD laser in the *l-th* layer are given by:

$$\frac{dS^{l,m}}{dt} = \left(\Gamma_p v_g g_{GS}^{l,m} - \frac{1}{\tau_P}\right) S^{l,m} + \beta_{SP} \frac{N_{GS}^l}{\tau_{GS}^{spon}}$$
$$+ 2k_c \sqrt{S_{inj}^{l,m}(t)S^{l,m}(t)} \cos\phi^{l,m}(t)$$
$$+ 2k_c \sqrt{R_{ext}^l S^{l,m}(t-\tau_l)S^{l,m}(t)} \cos\left[\phi_0 + \phi^{l,m}(t) - \phi^{l,m}(t-\tau_l)\right] \quad (5)$$

$$\frac{d\phi^{l,m}}{dt} = \frac{1}{2}\Gamma_P v_g \left( g_{GS}^{l,m}\kappa_{GS} + g_{ES}^{l,m}\kappa_{ES} + g_{RS}^{l,m}\kappa_{RS} \right)$$
$$-k_c \sqrt{\frac{S_{inj}^{l,m}(t)}{S^{l,m}(t)}} \sin\phi^{l,m}(t) - \Delta\omega_{inj}^{l,m}$$
$$-k_c \sqrt{\frac{R_{ext}^l S^{l,m}(t-\tau_l)}{S^{l,m}(t)}} \sin\left[\phi_0 + \phi^{l,m}(t) - \phi^{l,m}(t-\tau_l)\right] \quad (6)$$

where $\beta_{sp}$ is the spontaneous emission factor. For the first hidden layer, $S_{inj}^{l,m}$ is the photon number of the ML modulated by the input signal. For other hidden layers, the injected photon number is $S_{inj}^{l,m}(t) = A_{inj}^l S^{l-1,m}(t)$ with $A_{inj}^l$ being an attenuation factor. This attenuation factor roughly equals to the injection ratio, which is the power ratio of the laser in the $(l-1)$-th layer to the laser in the $l$-th layer. $\Delta\omega_{inj}^{l,m}$ is the detuning frequency between the two lasers. $R_{ext}^l$ is the feedback power ratio, $\tau_l$ is the feedback delay time of the $l$-th layer, and $\phi_0$ is the initial feedback phase. $k_c$ is the coupling coefficient of the injection or feedback light into the laser cavity. $g_{RS}^{l,ES,}$ is the gain of the ES and the RS. $\kappa_{ES,RS}^{GS,}$ is a contribution coefficient of each state to the linewidth broadening factor (LBF). In order to simplify the discussion, all the SLs in the deep PRC are assumed to be identical, although different SLs are helpful to enrich the neuron dynamics and thereby to improve the performance.[36] In addition, both the optical injection condition and the optical feedback condition for all the hidden layers are kept the same as well. Besides, the model does not take into account the optical noise in the system.

**TABLE I.** Main parameters of the laser and the deep PRC.[35]

| Parameter | Symbol | Value |
|---|---|---|
| RS spontaneous lifetime | $\tau_{RS}^{spon}$ | 0.5 ns |
| ES spontaneous lifetime | $\tau_{ES}^{spon}$ | 0.5 ns |
| GS spontaneous lifetime | $\tau_{GS}^{spon}$ | 1.2 ns |
| Spontaneous emission factor | $\beta_{sp}$ | $1.0\times10^{-4}$ |
| Photon lifetime | $\tau_p$ | 4.1 ps |
| GS contribution to LBF | $\kappa_{GS}$ | 0.25 |
| ES contribution to LBF | $\kappa_{ES}$ | 0.12 |
| RS contribution to LBF | $\kappa_{RS}$ | 0.04 |
| Optical confinement factor | $\Gamma_p$ | 0.06 |
| Coupling coefficient | $k_c$ | $1.0\times10^{11}$ /s |
| Attenuation factor | $A_{inj}$ | -5 dB |
| Detuning frequency | $\Delta\omega$ | 0 GHz |
| Feedback ratio | $R_e$ | -28 dB |
| Initial feedback phase | $\phi_0$ | 0 |
| Normalized delay time | $\tau/T_c$ | 1.5 |
| Neuron number per layer | $N_e$ | 50 |
| Neuron separation | $\theta$ | 10 ps |
| Hidden layer number | / | 3 |
| Clock cycle | $T_c$ | 500 ps |

## III. SIMULATION RESULTS

The main parameters of the laser and the deep PRC used in the simulation are listed in Table I, unless stated otherwise. Detailed parameters of the laser are in Ref. 35. The main parameters of the optical transmission system under study are listed in Table II, unless stated otherwise. The amplified spontaneous emission noise from the EDFA is included in the transmission system, while the phase noise of the laser source is not considered. The SL under study exhibits a lasing threshold of $I_{th}$ = 50 mA. The laser is biased at 1.6 × $I_{th}$, leading to a photon number of 9.2 × 10$^4$. The resonance frequency of the laser is 2.8 GHz, resulting in a characteristic time of 360 ps. The neuron interval is fixed at $\theta$ = 10 ps, which is much smaller than the characteristic time to ensure the instantaneous state of the deep PRC system. The feedback delay time $\tau$ is asynchronous with the clock cycle $T_c$, and normalized delay time is set at $\tau/T_c$ = 1.5.[51] The default depth of the PRC is 3, and the default neuron number in each hidden layer is $N_e$ = 50. In the optical transmission system, the symbol format is DP 16-QAM, and the symbol rate is fixed at 30 GBaud, corresponding to a bit rate of 240 Gbps. The default launch power of the transmitter is 12 dBm, and the transmission distance is 50 km. The transmission system is simulated using an open-source Python tool of Intelligent Fiber Transmission Simulation (IFTS).[52] This software supports various signal modulation formats and multiple DSP functions for the signal recovery.

**TABLE II.** Main parameters of the transmission system.[51]

| Parameter | Value |
|---|---|
| Carrier wave wavelength | 1550 nm |
| Fiber loss coefficient | 0.2 dB/km |
| Fiber dispersion parameter | 16.75 ps/nm/km |
| Fiber nonlinear coefficient | 1.32 W$^{-1}$ km$^{-1}$ |
| Polarization mode dispersion | 0 ps/$\sqrt{km}$ |
| EDFA noise figure | 5 dB |
| OSNR | 31.5 dB |
| Modulation format | DP 16-QAM |
| Symbol rate | 30 GBaud |
| Bit rate | 240 Gbps |
| Transmitter launch power | 12 dBm |
| Transmission distance | 50 km |
| Equalizer tap number | 7 |

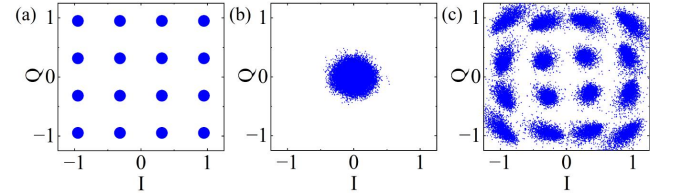

**FIG. 2.** Simulated constellation of 16-QAM signals (a) at the transmitter, (b) after the optical fiber link, and (c) after the DSP.

Figure 2 illustrates the simulated 16-QAM signals of x-polarization at the transmitter (Fig. 2(a)), after the optical fiber link (Fig. 2(b)), and after the DSP module (Fig. 2(c)), respectively. It is shown that the 16-QAM signal after the DSP processing remains significantly distorted, due to the impairment of fiber nonlinearity. This distorted signal is then fed into the deep PRC for the nonlinear equalization. 64,880 random symbols are used to train the deep PRC by the algorithm of ridge regression, while 32,440 random symbols are used to test the performance. The performance of nonlinear equalization is quantified by the Q factor, which is determined by the BER through the following formula:[53]

$$Q = 20\log_{10}(\sqrt{2} \cdot erfcinv(2 \cdot \mathrm{BER})) \quad (7)$$

where *erfcinv* is the inverse complementary error function. Generally, the Q factor increases with decreasing BER, and a larger Q factor suggests a better equalization performance.

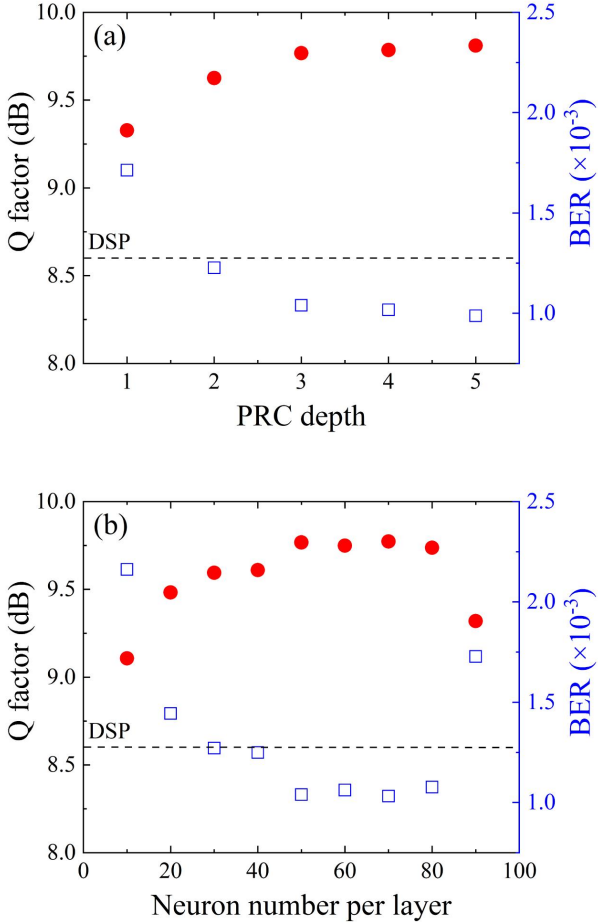

**FIG. 3.** Impacts of (a) the PRC depth and (b) the neuron number on the Q factor (dots) and the BER (squares). Dashed line denotes the Q factor obtained by the DSP alone. The other parameters of the PRC are listed in Table I.

We firstly investigate the impact of the PRC depth on the performance of nonlinear equalization in Fig. 3(a). Without using the PRC, the sole DSP provides a Q factor of 8.62 dB (dashed line), corresponding to a BER of 3.5×10$^{-3}$. When applying the PRC with one hidden layer, the Q factor is raised to 9.33 dB. Increasing the PRC depth raises the Q factor, which becomes 9.77 dB at the depth of 3. However, the performance is almost saturated when further increasing the depth. Figure 3(b) shows that the Q factor increases from 9.11 dB for the neuron number of $N_e = 10$ up to 9.77 dB for $N_e = 50$. On the other hand, the Q factor has little change in the range of $N_e = 50$ to 80. It decreases at $N_e = 90$ due to the overfitting problem.

Fixing the depth of the PRC at 3 and the neuron number at 50, Fig. 4 investigates the effects of the optical feedback conditions and the optical injection conditions on the performance of nonlinear equalization. The Q factor in Fig. 4(a) firstly goes up with the normalized delay time $\tau/T_c$, and then reaches the maximum value of 9.77 dB at $\tau/T_c = 1.5$. This suggests the optimal feedback delay time $\tau$ is asynchronous with the clock cycle $T_c$.[51] In comparison with the common synchronous architecture, this asynchronous one is helpful to enrich the interconnections of neurons and hence to raise the memory capacity.[51,54] The critical feedback level of the SL (without optical injection) is -11 dB, beyond which the laser becomes unstable. Figure 4(b) shows that the Q factor slightly increases from 9.64 dB at $R_{ext} = -46$ dB to 9.77 dB for feedback ratios ranging from -28 dB to -22 dB. The Q factor declines for feedback ratios beyond -20 dB, which is still far below the critical feedback level of -11 dB. Therefore, the PRC does not necessarily achieve the best performance at the boundary of stability.[55] On the other hand, the optical injection substantially alters the resonance frequency and the damping factor of semiconductor lasers.[56] At the injection ratio of -5 dB, the stable locking regime of the SL (without optical feedback) is bounded by the saddle-node bifurcation at -14 GHz and the Hopf bifurcation at +3.5 GHz. Figure 4(c) shows that the Q factor firstly rises with increasing attenuation factor, and reaches the maximum value of 9.77 dB at $A_{inj} = -5$ dB, beyond which the Q factor declines. In addition, through varying the detuning frequency from the side of the saddle-node bifurcation to the side of the Hopf bifurcation in Fig. 4(d), the optimal deep PRC performance is achieved at 0 GHz, which is more close to the Hopf bifurcation. This is because the optical injection with a detuning frequency in the vicinity of the Hopf bifurcation reduces the damping factor whereas enhances the resonance frequency of the slave laser, which leads to rich neuron dynamics.[57]

## IV. DISCUSSION

In the above section, we have obtained the optimal operation conditions (shown in Table I) for the nonlinear equalization of the 16-QAM signal with the launch power of 12 dBm and the transmission distance of 50 km. The optimal Q factor of the deep PRC is 9.77 dB, which is 1.15 dB larger than that of the DSP. This means the Q factor gain of the deep PRC is 1.15 dB. In comparison, a three-layer MLP (with 50 neurons per layer) exhibits a Q factor of 10.09 dB, while a one-layer bidirectional LSTM (with 50 hidden units) exhibits a Q factor of 10.08 dB. Both values are about 0.3 dB higher than that of the deep PRC. However, we can not simply conclude that the deep PRC performance is worse or better than the neural networks in this work. On one hand, the operation parameters of the deep PRC are not fully optimized. On the other hand, the deep PRC is an analog system, and hence the complexity analysis commonly used in the digital neural networks is not suitable for the PRC anymore.

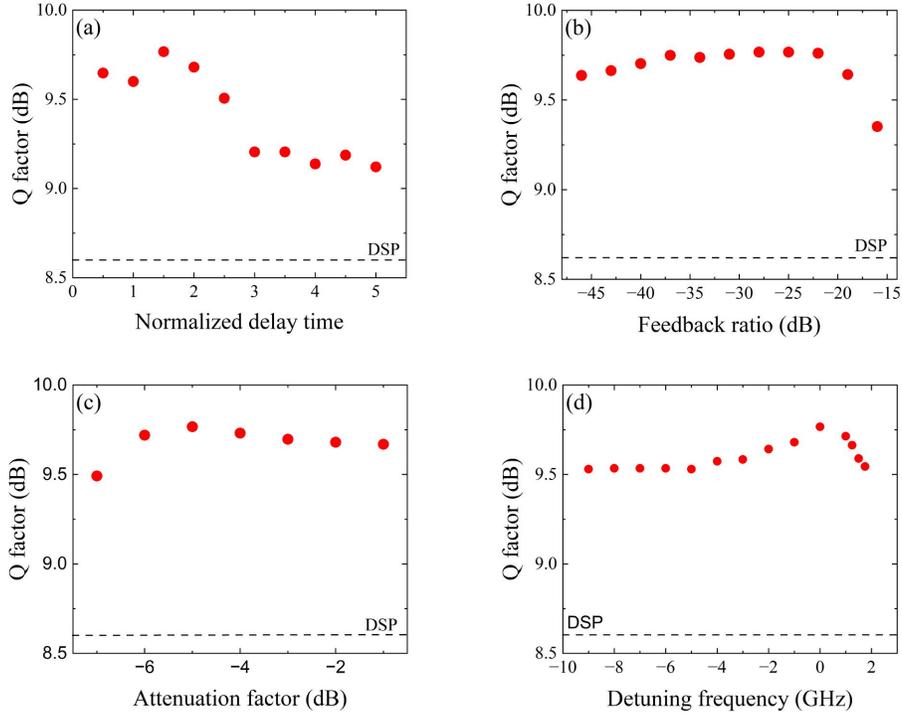

**FIG. 4.** Effects of (a) the delay time, (b) the feedback ratio, (c) the attenuation factor, and (d) the detuning frequency on the Q factor. The depth of the PRC is 3, and the neuron number is 50 in each layer. The other parameters of the PRC are listed in Table I.

Now, we apply the deep PRC with the same set of operation parameters in Table I to solve the nonlinear equalization at different launch powers and at different transmission distances, without further optimization. Figure 5(a) shows that the Q factors of the DSP (squares) and the deep PRC (dots) decline with increasing launch power from 10 dBm to 15 dBm, due to the increased fiber nonlinearity. In addition, Fig. 5(a) unveils that the Q factors of the deep PRC are always larger than that of the DSP, although the operation parameters of the deep PRC are not optimized at each corresponding launch power. Interestingly, the Q factor gain (triangles) generally rises with the launch power, from 1.01 dB at 10 dBm up to 1.59 dB at 15 dBm. Therefore, the deep PRC has strong capability in compensating for the nonlinearity impairment of 16-QAM signals. It is remarked that Fig. 5(a) does not show the performance below the launch power of 10 dBm. This is because the BER of the PRC below 10 dBm is smaller than $7.71 \times 10^{-6}$, which is beyond the resolution of tested number of symbols (32,440). Figure 5(b) and Fig. 5(c) illustrate the constellation diagrams of the 16-QAM signal with the nonlinear equalization of the deep PRC (red) at 10 dBm and at 12 dBm, respectively. It is shown that the deep PRC obviously improves the quality of the constellation diagram, where the data points are more concentrated than that of the DSP (blue).

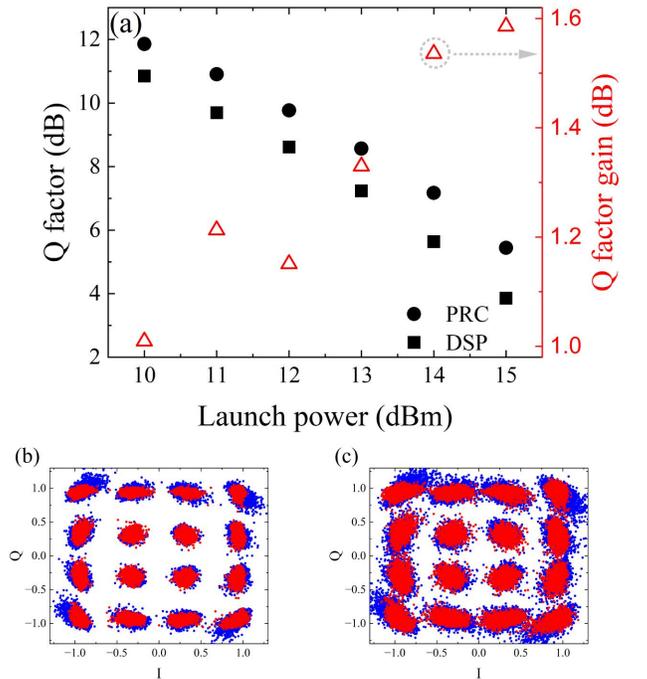

**FIG. 5.** (a) Q factor (dots) and Q factor gain (triangles) of the deep PRC versus the launch power of the transmitter. The squares stand for the Q factor of the DSP. Constellation diagrams of the 16-QAM signals with (red)

and without (blue) the nonlinear equalization of the deep PRC for launch powers of (b) 10 dBm and (c) 12 dBm. The transmission distance is 50 km, and the other parameters of the PRC are listed in Table I.

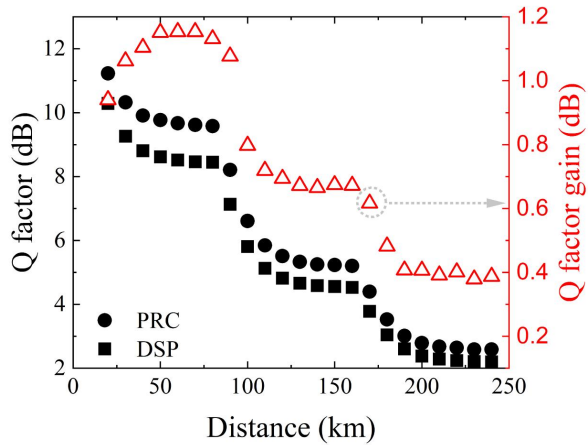

**FIG. 6.** Q factor (dots) and Q factor gain (triangles) of the deep PRC versus the transmission distance. The squares stand for the Q factor of the DSP. The depth of the PRC is 3, and the neuron number is 50 in each layer. The launch power is 12 dBm, and the other parameters of the PRC are listed in Table I.

Figure 6 shows the performance of the deep PRC on the nonlinear equalization of the 16-QAM signal at different transmission distances. Due to the transmission loss, the optical signal is amplified by an EDFA with a noise figure of 5 dB for every fiber span of 80 km long. The EDFA raises the optical signal power whereas reduces the signal-to-noise ratio (SNR), due to the addition of the amplified spontaneous emission noise.[58] It is shown that the Q factors of the DSP (squares) and the deep PRC (dots) decline with increasing transmission distance, due to the raised fiber nonlinearity and the reduced SNR. At the beginning of every fiber span (90 km and 170 km), the Q factor drops sharply, which is mainly attributed to the addition of the EDFA noise.[59] With increasing transmission distance, the Q factor gain firstly rises to the maximum value of 1.15 dB for ranges of 50 to 70 km, and then reduces down to 0.41 dB at 200 km. Surprisingly, the best equalization performance is not achieved at the shortest distance of 20 km, but at the range of 50 to 70 km. This is because the operation conditions of the deep PRC are optimized at 50 km rather than 20 km. This behavior of the Q factor gain suggests that it is necessary to optimize the operation parameters of deep PRC for every transmission distance to achieve the best signal recovery performance.

Future work will investigate the performance of the deep PRC in compensating for the nonlinearity of multi-channel WDM communication systems, where the inter-channel nonlinear effects play a crucial role.[60] The equalization of this inter-channel nonlinearity is very challenging for both DSP algorithms and the digital neural networks.[60,61] The proposed PRC architecture is able to deal with multiple channels in parallel, through deploying multiple longitudinal modes of the FP laser in Fig. 1(c).[42,45] Each channel of the 16-QAM signal consumes four modes, considering the two I/Q components and the two x/y polarizations. In addition, both the noise of the optical transmission system and the noise of the deep PRC system can degrade the nonlinear equalization performance of the 16-QAM signals,[42] which will be systematically investigated in the future work.

In comparison with the DSP counterpart, the power efficiency of the optical computing scheme is expected to be one to two orders of magnitude higher, and the latency is two to three orders of magnitude lower.[62,63,64] However, the footprint of optical computing chips is usually tens of mm$^2$, and hence is generally larger than the digital ones.

## V. CONCLUSION

In summary, we have proposed a deep PRC for the nonlinear equalization of 16-QAM signals with a transmission rate of 240 Gbps. The PRC relies on the cascading injection-locked FP lasers to achieve the deep architecture, and the wavelength multiplexing of FP modes to realize the parallel structure. It is found that the deep PRC exhibits strong capability in compensating for the impairment of fiber Kerr nonlinearity of 16-QAM signals. At a transmission distance of 50 km, the Q factor gain of the deep PRC is more than 1 dB, for high launch powers ranging from 10 dBm up to 15 dBm. Meanwhile, at a launch power of 12 dBm, the Q factor gain of the deep PRC is larger than 0.4 dB for transmission distances up to 200 km. Future work will optimize the operation parameters of the deep PRC layer by layer, instead of using identical parameters. In addition, we will demonstrate the corresponding nonlinear equalization performance of the deep PRC in experiment.

## ACKNOWLEDGMENTS

This work was funded by Science and Technology Commission of Shanghai Municipality (24TS1401500, 24JD1402400), and by National Natural Science Foundation of China (62475152, 62025503).

## AUTHOR DECLARATIONS

### Conflict of Interest

The authors have no conflicts to disclose.

### Author Contributions

Rui-Qian Li, Yi-Wei Shen and Zekun Niu contributed equally to this work.

**Rui-Qian Li**: Data curation (equal); Investigation (equal); Validation (equal); Visualization (equal); Writing-original draft (equal). **Yi-Wei Shen**: Data curation (equal); Investigation (equal); Validation (equal); Visualization (equal); Writing-original draft (equal). **Zekun Niu:** Data curation (equal); Investigation (equal); Validation (equal); Visualization (equal); Writing-original draft (equal). **Guozhi Xu**: Data curation (equal); Investigation (equal); Validation (equal). **Jingyi Yu**: Supervision (equal); Writing - review & editing (equal). **Xuming He**: Supervision (equal); Writing - review & editing (equal). **Lilin Yi**: Supervision (equal); Writing - review & editing (equal). **Cheng Wang**: Methodology (lead); Project administration (lead); Supervision (lead); Funding acquisition (lead); Writing - review & editing (lead).

## DATA AVAILABILITY

The data that support the findings of this study are openly available in https://zenodo.org/records/14506266.[65]